\documentclass[aps,twocolumn,showpacs]{revtex4}
\usepackage{graphicx}

\begin{document}
\title{Breakdown electron-hole symmetry in graphene structure \\ 
with a semiconductor gate}
\author{F. T. Vasko}
\email{ftvasko@yahoo.com}
\affiliation{QK Applications,  3707 El Camino Real, San Francisco, CA 94033, USA}
\date{\today}

\begin{abstract}
The electron-hole symmetry in the structure ``graphene - insulating substrate -semiconductor gate'' is violated due to an asymmetrical drop of potential in the semiconductor gate under positive or negative biases. The gate voltage dependencies of concentration and conductivity are calculated for the case of SiO$_2$ substrate placed over low- (moderate-) doped $p$-Si. Similar dependencies of the optical conductivity are analyzed for the case of high-$\kappa$ substrates (AlN, Al$_2$O$_3$, HfO$_2$, and ZrO$_2$). The comparison of our results with experimental data shows a good agreement for both cases.
\end{abstract}


\maketitle
The energy spectrum and scattering mechanisms in graphene are symmetric with respect to electron-hole replacement, so that the transport phenomena are identical for the electron and hole doping cases, see reviews 1 and 2. According to the theoretical analysis, as well as the transport and optical measurements (see review \cite{3} and references therein), the Coulomb renormalization in graphene leads to weak ($\sim$10 \% for typical parameters) asymmetry of electron and hole responses. These results were obtained for the electrostatically doped graphene structures with metallic (or heavily-doped semiconductor) gates. The sheet concentration of carriers, $n_g$, is determined by the gate voltage $V_g$ according to the plane capacitor formula 
%1
\begin{equation}
n_g  =\alpha (V_g -V_0) , ~~~~ \alpha\approx\frac{\epsilon_s}{4\pi |e|d}  ,
\end{equation}
where $\epsilon_s$ is the dielectric permittivity of substrate of thickness $d$, see Fig. 1a, and voltage $V_0$ corresponds to the electroneutrallity point; below we suppose $V_0\to 0$. Thus, transport phenomena should be symmetric with respect to the sigh flip of $V_g$ (electron-hole symmetry). But this symmetry appears to be violated for the structures with low- (moderate-) doped  semiconductor gates. It is due to differences in the drop of potential and charge distributions for positive or negative biases, when the depleted or heavily-doped regions appear, see Figs. 1b and 1c, respectively. This simple mechanism was not analyzed yet and it is timely to re-examine Eq. (1) for the case of low-doped semiconductor gates, which were used in several experiments. \cite{4,5,6}
%f1
\begin{figure}[tbp]
\begin{center}
\includegraphics[scale=0.6]{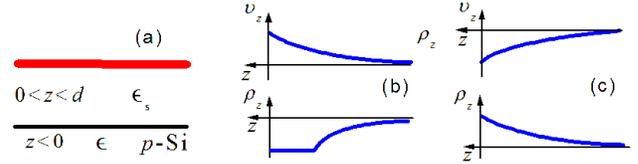}
\end{center}
\addvspace{-0.5cm}
\caption{(a) Structure schematic for back-gated graphene (grey) with dielectric
substrate of thickness $d$ and permittivity $\epsilon_s$ over $p$-doped Si of permittivity $\epsilon$. (b) and (c) Potential (upper) and charge (lower) distributions along semiconductor gate for the electron and hole doped Si cases, respectively. }
\end{figure}

In this letter we solve the electrostatic problem for graphene structures with thick SiO$_2$ or thin high-$\kappa$ (AlN, Al$_2$O$_3$, HfO$_2$, and ZrO$_2$) substrates placed over $p$-Si of different doping levels. The gate voltage dependencies of the concentration and the dc conductivity as well as the interband optical conductivity are calculated and comparisons with the experimental data \cite{4,5,6} are performed.

The distribution of the electrostatic potential $\varphi_z$ across the structure under consideration (along $z$-axis, $z\leq d$) is governed by the Poisson equation
%2
\begin{equation}
\frac{d}{dz}\left( \epsilon_z \frac{d\varphi_z}{dz}\right) =-4\pi\rho_z ,~~~~ \epsilon_z =\left\{ \begin{array}{*{20}c} {\epsilon_s ,} & {0 < z < d}  \\
   {\epsilon ,} & {z < 0}  \\
\end{array} \right. ,
\end{equation}
where $\epsilon_z$ stands for the dielectric permittivity and $\rho_z$ is the charge density in graphene (at $z\to d$) and in Si gate (at $z<$0). Using the Gauss theorem at $z\to d-0$ and connecting the derivatives of $\varphi_z$ at $z\to 0$ one obtains the boundary conditions
 %3
\begin{equation}
\left. \frac{d\varphi_z}{dz} \right|_{z=d-0} =\frac{4\pi}{\epsilon }\sigma_g ,  ~~~~~~  \epsilon_z \left. \frac{d\varphi_z}{dz} \right|_{-0}^{+0} =0 .  
\end{equation}
Here $\sigma_g =\int_{d-0}^{d+0}dz\rho_z =\mp |e|n_g$ is the charge density induced in graphene which is written through $n_g$ and $-$ or $+$ correspond to the electron or hole doping. The linear potential distribution $\varphi_z =\varphi_{z=0} +4\pi\sigma_g z/\epsilon_s$ takes place in the substrate region $d>z>0$, where $\rho_z =0$. In the gate region $z<0$ Eq. (2) is transformed into the second-order differential equation
%4
\begin{equation}
\frac{d^2 \varphi_z}{dz^2} = \frac{{4\pi |e|}}{\epsilon }\left\{ {\begin{array}{*{20}c}
{n_A  - n_{hz} ,} & ~(p{\rm -Si})  \\
   {n_{ez}  - n_D ,} & ~(n{\rm -Si}) 
\end{array}} \right.  ,
\end{equation}
which is written for $p$-Si ($n$-Si) through the differences between acceptor (donor) and hole (electron) concentrations, $n_A$ ($n_D$) and $n_{hz}$ ($n_{ez}$). The boundary conditions for Eq. (4) are $\varphi_{z\to -\infty}=0$ and the right Eq. (3), which determines an electric field at $z\to -0$.

Below, we restrict ourselves by the case of $p$-doped Si, where the non-degenerate hole approximation $n_{hz}=n_A\exp (e\varphi_z /T)$ is valid for 
$n_A\leq 5\times 10^{18}$ cm$^{-3}$ at room temperature $T=$300 K. For higher $n_A$ or for the $n$-doping case, $n_{hz}$ or $n_{ez}$ depend weakly on $\varphi_z$ and deviations from Eq. (1) are weak. Since Eq. (4) does not depend on $z$ explicitly, we obtain the first-order nonlinear equation
%5
\begin{equation}
\left( \frac{e}{T}\frac{d\varphi _z }{dz} \right)^2  =\frac{2}{z_T^2}\left( {e^{e\varphi _z /T}-\frac{{e\varphi _z }}{T} - 1} \right) ,
\end{equation}
where the characteristic length $z_T$ is determined by the relation  $(4\pi e^2 /\epsilon T)n_A z_T^2 =1$. Further integration gives the potential and charge distributions, which are schematically shown in Figs. 1b and 1c, and $\varphi_{z=0}$ is connected with $n_g$ through the boundary condition at $z\to -0$:
%6
\begin{equation}
e^{\upsilon_0}-\upsilon_0 -1 = \frac{1}{2}\left( {\frac{{n_g }}{{n_T }}} \right)^2 , ~~~ \upsilon_0\equiv\frac{e\varphi_{z = 0}}{T}  .
\end{equation}
Here we introduce the characteristic 2D concentration $n_T =n_A z_T\propto\sqrt{n_A T}$ which is about $0.8\times 10^{10}$ cm$^{-2}$ if $n_A =10^{15}$ cm$^{-2}$ and $T=$300 K. Neglecting the low-doped region $n_g\leq n_T$, we obtain the simple relation between $V_g$ and $n_g$:
%7
\begin{equation}
\alpha V_g\simeq\mp n_g +n_T \frac{\epsilon_s z_T}{\epsilon d}\left\{ \begin{array}{*{20}c}
   \left( {n_g /n_T} \right)^2 /2 , & {\rm (h)}  \\
	-2\ln \left( {n_g /\sqrt 2 n_T } \right), & {\rm (e)}  
\end{array} \right.  ,
\end{equation}
which generalize the standard plane capacitor formula given by Eq. (1). For the electron-doping case ($V_g >0$), both $\propto n_g$ and $\propto n_g^2$ contributions are essential depending on the dimensionless factor $\epsilon_s z_T /\epsilon d$. By contrast, for the hole-doping case ($V_g <0$) the $\ln$-correction can be only detected in the low-doped graphene. Note, that there is no temperature dependency for $V_g <0$ and these dependencies are weak (appears in $\ln$-correction) for $V_g >0$.
%f2
\begin{figure}[tbp]
\begin{center}
\includegraphics[scale=0.6]{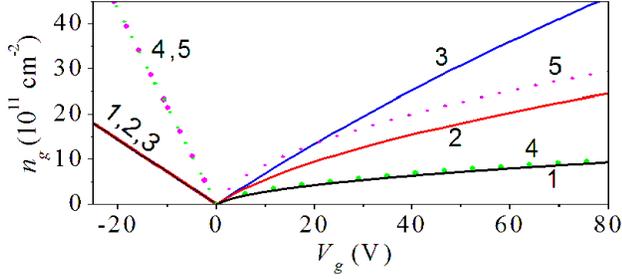}
\end{center}
\addvspace{-0.5 cm}
\caption{Concentration of carriers $n_g$ (electrons if $V_g >0$ or holes if $V_g <0$) versus gate voltage $V_g$ for graphene on SiO$_2$ substrate of 300 nm or 100 nm width (solid or dotted curves) placed over $p$-Si gate with doping levels: $n_A =10^{15}$ cm$^{-3}$ (1,4), $10^{16}$ cm$^{-3}$ (2,5), and $10^{17}$ cm$^{-3}$ (3).  }
\end{figure}

Fig.2 shows the carriers concentrations versus $V_g$ in graphene placed over the SiO$_2$/$p$-Si structure depending on doping level and substrate thickness, $n_A$ and $d$. These dependencies are substantially different for $e$- and $h$-doping regimes. If $V_g<0$, the plane capacitor approximation (1), with the $n_A$-independent concentration $n_g\propto d^{-1}$, is valid. If $V_g>0$, the dependency $n_g (V_g )$ varies from the square-root to linear function and the electron-hole symmetry is restored if $n_A$ exceeds $10^{17}$ cm$^{-3}$ at $d=$300 nm. For thiner substrates ($d=$100 nm is shown), the symmetry is restored at higher concentrations. The square-root dependency for low-doped $p$-Si appears due to the depleted region of thickness $d_{\rm depl}$ (see Fig. 1b), which is determined by the charge neutrality condition, $n_s\simeq n_A d_{\rm depl}$. Using Eq. (1) with $d\to d+d_{\rm depl}$, one obtains $n_s\propto\sqrt{V_g}$ for the case $d_{\rm depl}\gg d$.
%f3
\begin{figure}[tbp]
\begin{center}
\includegraphics[scale=0.6]{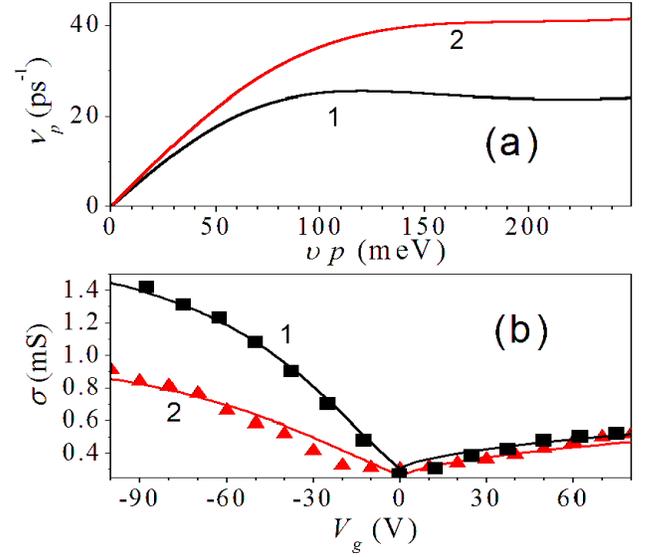}
\end{center}
\addvspace{-0.5 cm}
\caption{(a) Fitting of experimental data from Refs. 4 and 5 marked as 1 and 2: (a) momentum relaxation rate $\nu_p$ versus energy $vp$ and (b) conductivity $\sigma$ versus  $V_g$. Squares and triangles are experimental points from Refs. 4 and 5, respectively. }
\end{figure}

The asymmetrical electrostatic doping of graphene under sign-flip of $V_g$ leads to similar changes of the gate voltage dependencies of static conductivity, $\sigma (V_g )$. These dependencies in structures placed on $p$-Si with $n_A\simeq 10^{15}$ cm$^{-3}$ were measured \cite{4,5} under investigations of the Drude and interband absorption in large-area samples. In order to fit $\sigma (V_g)$ we use the phenomenological momentum relaxation rates $\nu_p$ suggested in Ref. 7 with the parameters taken from the experimental data of Refs. 4 and 5 for the hole doping cases $V_g <0$, when Eq. (1) is valid, see Fig. 2. Using the relaxation rates shown in Fig. 3a and the dependencies $n(V_g)$ for $V_g >0$, we obtain an excellent agreement between $\sigma (V_g )$ and the results of Refs. 4 and 5, as it is shown in Fig. 3b. The asymmetry of $\sigma (V_g)$ is more pronounced than $n(V_g)$ (c.f. Figs. 2 and 3b) because of an additional energy dependency of $\nu_p$.
%f4
\begin{figure}[ht]
\begin{center}
\includegraphics[scale=0.6]{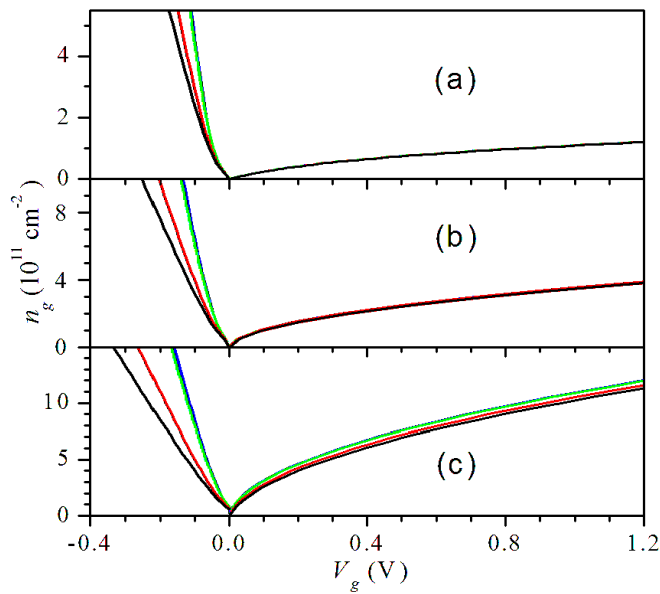}
\end{center}
\addvspace{-0.5 cm}
\caption{(a) Concentration $n_g$ versus $V_g$ for graphene on AlN (black), Al$_2$O$_3$ (red), HfO$_2$ (green), and ZrO$_2$ (blue) substrates of width 10 nm placed over $p$-Si of doping level $10^{15}$ cm$^{-2}$. (b) and (c) The same for $p$-Si of doping levels $10^{16}$ cm$^{-2}$ and $10^{17}$ cm$^{-2}$, respectively. }
\end{figure}

Further, we turn to consideration of the structure with a thin high-$\kappa$ substrates placed over $p$-Si. For these structures, Eq. (7) contains the parameter $\epsilon_s z_T /\epsilon d\gg 1$ and a deviation from the standard formula (1) increases. Similar to Fig. 2, the dependencies of concentrations $n_s$ on $V_g$ are shown in Fig. 4 for AlN, Al$_2$O$_3$, HfO$_2$, and ZrO$_2$ substrates \cite{8} of width 10 nm at different doping levels. Here the square-root behavior  of $n(V_g)$, which is weakly dependent on $\epsilon_s$, takes place at $V_g >0$. For the hole doping regime, concentrations $n(V_g )$ deviate from the linear functions at low $|V_g |$, due to the $\ln$-contributions in Eq. (7), and curves are different for different substrates due to variations of $\epsilon_s$. For higher doping levels $n_A$, the induced concentration $n_g$ increases with $V_g$ faster, c.f. Figs. 4a-4c where $n_A$ varies from $10^{15}$ to $10^{17}$ cm$^{-3}$. The interval of $V_g$ applied is restricted by the breakdown condition $V_g /d <$5 - 10 MV/cm, \cite{9} so that we consider the region $|V_g |\leq$5 V.
%f5
\begin{figure}[tbp]
\begin{center}
\includegraphics[scale=0.6]{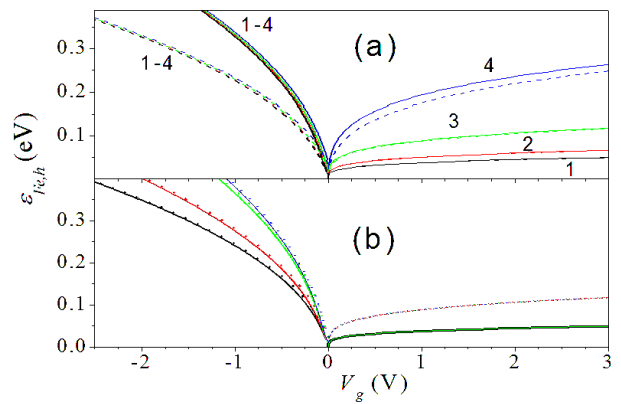}
\end{center}
\addvspace{-0.5 cm}
\caption{Electron and hole Fermi energies versus gate voltage $V_g$ for Al$_2$O$_3$ substrate of widths $d=$7 and 15 nm (solid and dashed curves) at $p$-Si doping levels 10$^{15}$, 10$^{16}$, 10$^{17}$, 10$^{18}$ cm$^{-3}$ marked 1 to 4, respectively. (b) The same for different substrates [AlN (black), Al$_2$O$_3$ (red), HfO$_2$ (green), and ZrO$_2$ (blue)] of widths $d=$10 nm at $p$-Si doping levels 10$^{15}$ and 10$^{17}$ cm$^{-3}$ (solid and dashed curves, respectively). }
\end{figure}

The essential asymmetry of electron-hole doping changes conditions for interband absorption because of a different Pauli blocking effect for $V_g >0$ and $<0$. At low temperatures (or for heavily-doped graphene) the absorption is blocked in the spectral region $\hbar\omega\leq 2\varepsilon_F$. Here the Fermi energies of electrons and holes $\varepsilon_F\propto\sqrt{n_g}$ depend on $V_g$ asymmetrically, see Figs. 2 and 4. So that, the threshold of absorption spectra for $V_g >0$ and $<0$ should be different: at fixed $\hbar\omega$, higher threshold bias is necessary for the electron doping regime, i.e. the asymmetric Pauli blocking takes place. In Fig. 5a we plot the electron and hole Fermi energies versus the gate voltage for structures with Al$_2$O$_3$ substrate of different widths placed over $p$-Si of different doping levels. Similarly to Fig. 4, there is a visible dependency on $n_A$ if $V_g >0$ while for $V_g <0$ the dependency of $\varepsilon_F$ on $d$ is only essential. The dependencies $\varepsilon_F (V_g )$ for structures with the all high-$\kappa$ substrates under consideration are shown in Fig. 5b for low- and heavily doped $p$-Si. Once again, $\varepsilon_F (V_g )$ is not dependent on $\epsilon_s$ at $V_g >0$ and the dependencies on $n_A$ are negligible at $V_g <0$. Note, that the threshold of absorption appears at low voltages for the hole doping case, $V_g\geq -1$ V if $\hbar\omega\leq$0.6 eV. For the electron doping case, there is no threshold up to the breakdown voltage, if $\hbar\omega\geq$0.3 eV.

An essential distinction between the spectral dependencies of interband absorption for the electron and hole doping regimes was reported in Refs. 4 and 5 for the structures of large-area graphene placed on SiO$_2$ substrate over low-doped $p$-i. These results are in a qualitative agreement with the asymmetrical dependencies of $\varepsilon_F (V_g )$ which can be obtained form Fig. 2. But a quantitative description of the dynamic conductivity $\sigma_\omega$ should take into account both the temperature broadening and the disorder effects, similar the consideration of the hole doping regimes performed recently, \cite{10} and it requires a special consideration. Here we consider the asymmetrical threshold of interband absorption in the graphene-based modulator with the Al$_2$O$_3$ substrate of width 7 nm and the waveguide designed for the telecommunication spectral range, $\hbar\omega\simeq$0.8 eV. \cite{6}  The depth of modulation, $\Delta T_\omega$, should be proportional to the carrier-induced contribution of the dynamic conductivity, $\Delta T_\omega\propto {\rm Re}\Delta\sigma_\omega $, where ${\rm Re}\Delta\sigma_\omega$ is given by \cite{10}
%8
\begin{equation}
{\rm Re}\Delta\sigma_\omega =\frac{\sigma_{\rm inter}}{\exp [(\hbar\omega  - 2\varepsilon_F)/2T]+1} , ~~~  \sigma_{\rm inter}=\frac{e^2}{4\hbar} .
\end{equation}
Here we consider the case $\varepsilon_F \gg T$, so that the chemical potential is replaced by $\varepsilon_F$. Using $\varepsilon_F (V_g )$ shown in Fig. 5, below we consider the dependency of $\Delta T_\omega\propto {\rm Re}\Delta\sigma_\omega$ on $V_g$ normalized to the unit amplitude of modulation.

The normalized conductivity $2{\rm Re}\sigma_\omega /\sigma_{\rm inter}$ is plotted in Fig. 6 for the conditions of Ref. 6 at different doping levels of $p$-Si. One can see, that the experimental data for the normalized dependency of $\Delta T_\omega$ on $V_g$ can be fitted at $n_A\simeq 3\times 10^{18}$ cm$^{-3}$ and $T=$300 K (a doping level is not given in \cite{6} and we use $n_A$ as a single fitting parameter). The thresholds of absorption are in good agreement with the measurements both for the electron and hole doping regimes but the temperature-induced smearing leads to more abrupt jumps in comparison to the experimental spectra. It is because a disorder contribution did not taken into account, see similar analysis of the interband absorption spectra \cite{4,5} in Ref. 10. In addition, the narrow waveguide structure forms a strip-like capacitor with the non-uniform charge distribution across the waveguide; \cite{11} this leads to an additional smearing effect. There is a weak doping dependency of $\Delta T_\omega$ at $V_g <0$ and at $V_g >0$ the modulation condition $\hbar\omega\sim 2\varepsilon_F$ satisfies if $n_A\geq 10^{18}$ cm$^{-3}$. The last estimate is valid for the in-plane homogeneous structure, while for a bus waveguide \cite{6} doping level should be lower. A complete description of the results of Ref. 6 requires to take into account both the disorder contributions and the waveguide geometry. 
%f6
\begin{figure}[tbp]
\begin{center}
\includegraphics[scale=0.6]{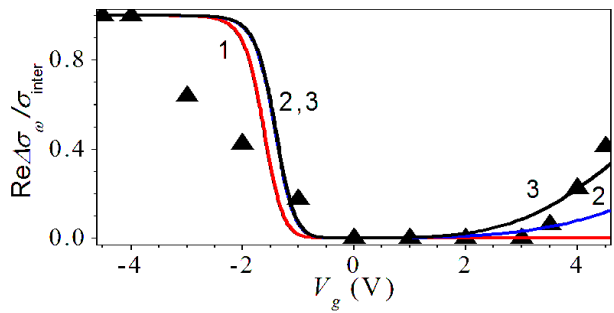}
\end{center}
\addvspace{-0.5 cm}
\caption{Dissipative interband conductivity ${\rm Re}\Delta\sigma_\omega$, normalized to $\sigma_{\rm inter}$, versus  $V_g$ for graphene on Al$_2$O$_3$ substrate at $p$-Si doping levels 10$^{15}$ (1), $2\times 10^{18}$ (2), and $3\times 10^{18}$ cm$^{-3}$ (3). Triangles show experimental data for $\Delta T_\omega$ from Ref. 6 which are normalized to the amplitude of modulation.  }
\end{figure}

Now we list the assumptions used in our calculations. The consideration of the non-degenerate holes is valid for any $n_A$ and any $V_g$ because of formation of the depletion region at $V_g >0$ and weak corrections to Eq. (1) at $V_g <0$. Apart from the mechanism under consideration, the Coulomb renormalization in $n$- or $p$-doped graphene  leads to an additional (up to 10 \%) asymmetry of response. \cite{3,12} Such a contribution can be neglected in the structures analyzed here (with a low-doped $p$-Si gate or a thin high-$\kappa$ substrate). But in typical samples (with the electron-hole asymmetry $\leq$10\%) both the Coulomb effect and the electrostatic contribution should be taken into account. This could result in decrease or increase of the electron-hole asymmetry. In the vicinity of the zero bias, $V_g\approx 0$, the long-range disorder \cite{13} should modify the effect under consideration.

Summarizing, the mechanism of the electron-hole symmetry breakdown, caused by different drops of potentials in low-doped Si for opposite signs of bias, have been analyzed and the  results are in agreement with the recent transport and optical measurements. These results are also important for device applications such as the field effect transistors \cite{14} or the  high-frequency multipliers. \cite{15}

\end{document}